\documentclass[11pt]{article}
\usepackage{fullpage}
\usepackage[dvips]{graphicx}
\usepackage{cite,mathtext} 
\usepackage{amssymb,amsmath}
\begin{document}

\newcommand{\fv}{\frac{1}{p_{xz}}}
\newlength{\pw}
\setlength{\pw}{0.5\textwidth}






\begin{center} \Large \bf The SPHINX spectrometer \end{center}

\begin{center} 
The SPHINX Collaboration \\ Yu.M.Antipov, A.V.Artamonov, V.A.Batarin,
V.A.Bezzubov, O.V.Eroshin, \\
S.V.Golovkin, Yu.P.Gorin, V.N.Govorun, A.N.Isaev, 
A.S.Konstantinov,\\A.P.Kozhevnikov, V.P.Kubarovsky, 
V.F.Kurshetsov, A.A.Kushnirenko, \\ L.G.Landsberg, V.M.Leontiev,
M.V.Medinskiy, V.A.Medovikov, V.V.Molchanov, \\
V.V.Morozova, V.A.Mukhin, D.I.Patalakha, S.V.Petrenko, A.I.Petrukhin,\\
V.I.Rykalin, V.A.Senko, N.A.Shalanda, A.N.Sytin,
V.S.Vaniev, \\ D.V.Vavilov, V.A.Victorov, V.I.Yakimchuk, S.A.Zimin\\
{\sl IHEP, Protvino, Russia}. \\
V.Z.Kolganov, G.S.Lomkatsi, A.F.Nilov, V.T.Smolyankin,\\
{\sl ITEP, Moscow, Russia}.
\end{center}

\begin{abstract}

The paper describes the SPHINX facility which includes a
wide-aperture magnetic spectrometer with scintillation counters and
hodoscopes, proportional chambers and drift tubes, multichannel
electromagnetic and hadron calorimeters, a guard system, a RICH
velocity spectrometer and a hodoscopical threshold Cherenkov detector
for the identification of charged secondary particles. The SPHINX
spectrometer, in its last modification, had the possibility to record
$\sim 3000$--$4000$ trigger events per  an accelerator burst. The
spectrometer was 
used during the last decade in experiments with the 70~GeV
proton beam of the  IHEP  accelerator U-70. 
 
\end{abstract}

\section{Introduction \label{se:intro}}


In 1989--1999 extensive studies of diffractive-like baryon production
and OZI suppressed reactions,  searches for exotic and cryptoexotic
pentaquark states and several other rare processes were carried out
in experiments with the SPHINX facility at the proton beam of
the IHEP accelerator with energy $E_p=70$~GeV. The SPHINX facility
was the multipurpose wide-aperture spectrometer with the possibility
of complete reconstruction of 
charged and neutral secondary particles. Over the time
several modifications of the SPHINX detector were made and all the
measurements can be divided into two stages:
\begin{itemize}
\item[a)] First generation experiments with the ``old'' setup,
performed in
1989--1994. The main results of these measurements were published
between 1994-2000 (see~\cite{3,4,5,6,7,8,9,10,11,12,13,14,15} and review
papers~\cite{2,1}).  The 
short description of the first version of the SPHINX detector can be
found in~\cite{3}.

\item[b)] Second generation experiments with the completely upgraded
SPHINX detector, which is described in this paper.
\end{itemize}
The ``old'' and upgraded SPHINX setup had the same general structure. However,
after the upgrade the facility was equipped with a new tracking system,
new hodoscopes, a hadron calorimeter and a modernized RICH velocity
spectrometer, new electronics, a new DAQ system and on-line computers. As a
result we obtained, practically, a new setup which was capable to record
an order of magnitude more data per spill.

\begin{figure}[hbt]
\includegraphics[width=1.00\textwidth]{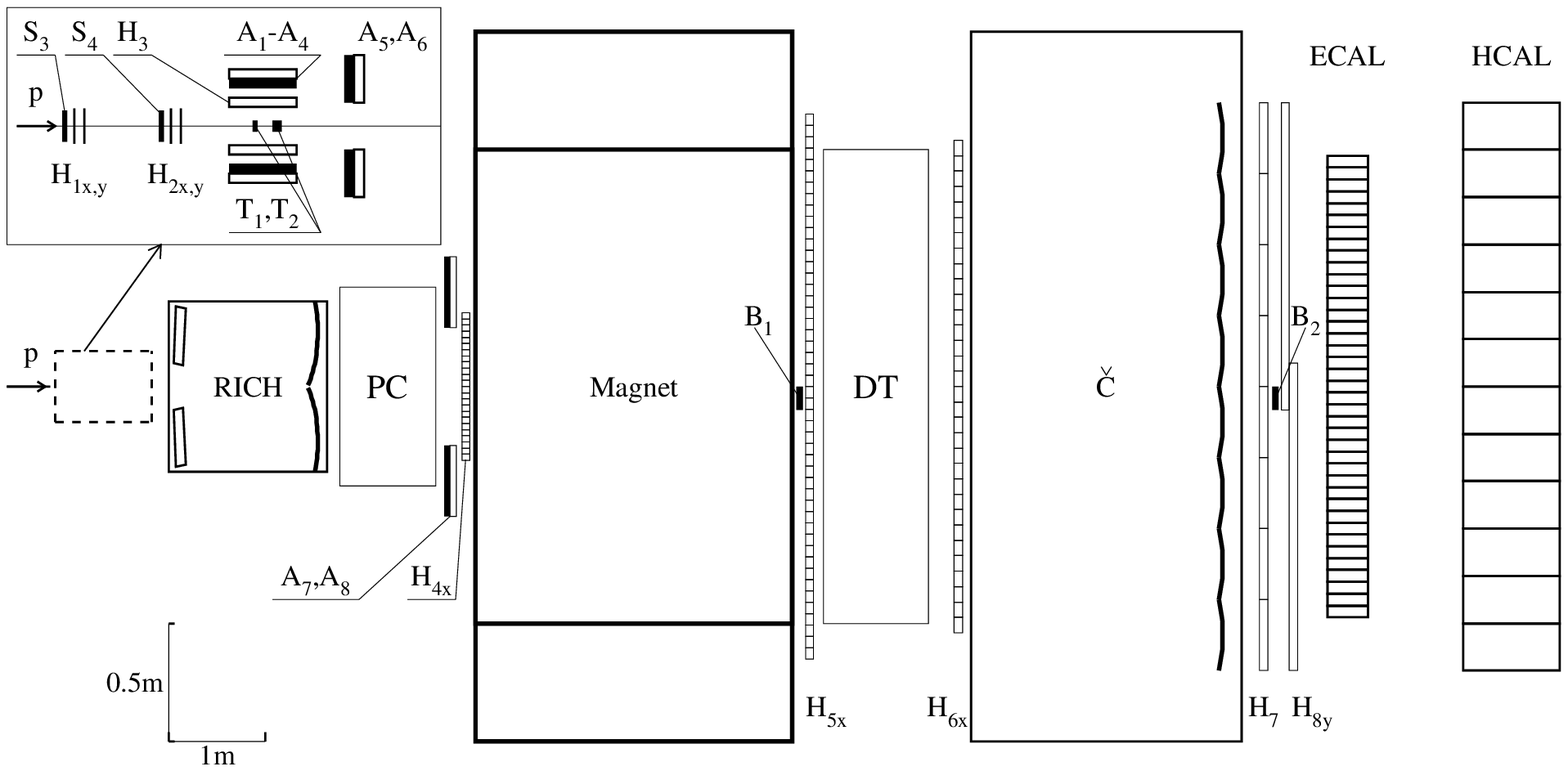}
\caption{%
Layout of the SPHINX spectrometer (plan view):
$\mathrm{S}_1$--$\mathrm{S}_4$~---
beam scintillation counters
(the very upstream counters $\mathrm{S}_1$ and $\mathrm{S}_2$ are not shown in this
figure);
$\mathrm{T}_1$,~$\mathrm{T}_2$~---
copper and carbon targets;
$\mathrm{A}_1$--$\mathrm{A}_8$~---
lead-scintillator veto counters;
$\mathrm{B}_1$--$\mathrm{B}_2$~---
veto counters to tag non-interacting primary beam particles;
$\mathrm{H}_{1X,Y}$, $\mathrm{H}_{2X,Y}$~--- beam hodoscopes;
$\mathrm{H}_3$~--- side hodoscope around the target;
$\mathrm{H}_{4X}$, $\mathrm{H}_{5X}$, $\mathrm{H}_{6X}$, $\mathrm{H}_{7}$, 
$\mathrm{H}_{8Y}$~--- trigger hodoscopes;
PC~--- proportional chambers;
DT~--- drift tubes;
RICH~--- velocity spectrometer with registration of the rings 
of Cherenkov radiation;
\v{C}~--- multichannel threshold Cherenkov counter;
ECAL~--- electromagnetic calorimeter;
HCAL~--- hadron calorimeter. 
}
\label{fig-sphinx21}
\end{figure}

The layout of the upgraded SPHINX detector
is  presented in Fig.~\ref{fig-sphinx21}. 
The right-handed $X,Y,Z$ coordinate system of the setup had $Z$-axis in 
the direction of the proton beam, vertical $Y$-axis and horizontal 
$X$-axis. The origin of the coordinate system was in the center of the
Magnet.

The main elements of the detector were as follows:
\begin{itemize}

\item[1.] The detectors of the primary proton beam~--- the
scintillation counters $S_1 - S_4$, $B_1, B_2$  and
the scintillator hodoscopes $H_{1X,Y}, H_{2X,Y}$. 

\item[2.] 
The targets $T_1$, $T_2$  and
the guard system --- the hodoscope $H_3$ and the lead-scintillator veto counters 
$A_1 - A_8$.


\item[3.] 
The wide-aperture magnetic spectrometer consisted of the magnet, the
block of proportional chambers~(PC), the assembly of drift tubes~(DT)
and the trigger hodoscopes $H_{4X}$, $H_{5X}$,
$H_{6X}$, $H_{7}$, $H_{8Y}$. 

\item[4.] 
Two Cherenkov counters (RICH, hodoscopical \v{C}) for secondary particles identification.

\item[5.] The multichannel lead glass electromagnetic calorimeter~(ECAL).

\item[6.]
The hadron calorimeter (HCAL). 

\item[7.]
Electronics, DAQ system and online computers.

\end{itemize}

During the runs from 1996 to 1999 with the upgraded SPHINX spectrometer
a flux of approximately 
$10^{12}$ protons has passed through the
targets and more than $10^9$ trigger events were
recorded on magnetic tapes.  
These statistics are now used to study numerous physical processes.
First results of these studies were published in~\cite{16,17,18}.

\section{Beam line and beam detectors}

The proton beam was transported to the setup by the multipurpose
beam-line channel~\cite{chan}. This,  300~m long, channel had quite a complex
structure utilizing 11 quadrapole, 12 kick, 2 correcting magnets and 3
types of collimators. The high intensity
($I\sim 10^{11}$--$10^{12}$~$p$/spill) of the primary beam of the U-70 accelerator
was reduced to $\sim 2$--$4\times10^6$~$p$/spill using a diffractive scattering in the thin targets
inside the channel. The beam was delivered to the target with
a negligible momentum spread, small space dimensions ($\sim 2\times
4$~mm$^2$) and a small angular divergence ($\sim 0.6$~mrad in each of $X,Y$-projections).

The intensity of the beam was measured by the  counters $S_1-S_4$.
The beam small cross section allowed the effective use of the $B_1,\ B_2$
counters as a beam killer telescope. 

All the beam counters were made of NE-102 scintillator and
were 5~mm thick. The $S_1$, $S_2$, $B_2$ counters were round with a diameter of 60~mm,
the $S_3$,
$S_4$, $B_1$ were rectangular with dimensions $30\times30$~mm$^2$. 
All the surfaces were polished and wrapped with aluminized
mylar. The light was transported to photo-multipliers (PMs) by plexiglas light-guides
glued to the counters. Fast FEU-87 PMs were used for the $S_1,\ S_2$ which
worked at higher intensity. For the other mentioned counters FEU-85 were used. Last
dinodes of all the PMs were powered from separate high voltage
sources to improve their stability. 

The position information of incident particles was measured with
the scintillator hodoscopes $H_{1X,Y}, H_{2X,Y}$. 

The $H_{1X,Y}$ were made of fiber bundles manucatured by KURARAY.
Each bungle consisted of
six layers of 
individually coated SCSF78M fibers, 0.5~mm in diameter each, glued
together. The non-transparent two layer coating served for reducing an
inter-fiber light interference. The hodoscope sensitive regions were
32~mm. 
The fibers were 350~mm long. They were divided by groups so
that each group corresponded to 2~mm in the direction perpendicular to
the beam. The ends were polished and each group was attached to the
individual PM FEU-85. Thus each hodoscope had 16 channels in total.

The $H_{2X,Y}$ were made of fiber bundles manufactured in
the IHEP (Protvino, Russia). The fibers had a single layer coating
and were much bigger in diameter (3.6~mm).
There were only two layers of fibers in the beam direction shifted in
respect to each other by half of the diameter. The hodoscope sensitive regions were
$\sim30$~mm. Each fiber
was attached to the individual PM FEU-85. Each hodoscope had 16 channel in total.

Signals from all the hodoscope PMs were transported to 
latch modules of registering electronics via 30~m long
$50\,\Omega$ coaxial cables. The PMs were powered by MEL high voltage sources
via resistor chains.

\section{Targets and guard system}
Two targets $T_1$ (Cu; 2.64~g/cm$^2$) and $T_2$ (C; 11.3~g/cm$^2$) were
separated by a distance of 25~cm and surrounded by a system of
counters. The system included the scintillator hodoscope $H_3$ and the veto counters
$A_1-A_4$, located around the targets, and $A_{5,6}$ 
(physically it was one counter seen by two PMs) which covered the
forward direction. Another part
of the veto system --- the counters $A_7,\ A_8$  were 
located further downstream, just before $H_{4X}$ hodoscope, at the
entrance to the magnet. The holes in
the counter $A_{5,6}$ and the distance between $A_7$, $A_8$ were
matched with the acceptance of the 
spectrometer. The $A_7$, $A_8$ veto counters are shown schematically in the
figure. In reality these two rectangular plates covered the entrance to the
magnet from the top and the bottom. 
The information from the $H_3$ and the veto system was used
to select exclusive reactions. 

The $H_3$ hodoscope consisted of 16 scintillators of $10\times 25\times500$~mm$^3$
each connected to individual PM FEU-85 by a plexiglas light-guide.

The $A_1-A_4$ counters --- led-scintillator sandwiches  were made of
five $200\times 500$~mm$^2$ sheets of each material, 5~mm thick for lead
and 10~mm thick for scintillator, put to an aluminum box. The scintillators were seen from one
end by FEU-30 PMs (one per the counter).
 
The $A_{5,6}$ and
$A_7,\ A_8$ counters were two layer
sandwiches made of one $500\times 500\times 10$~mm$^3$ lead and one $500\times
500\times 20$~mm$^3$ scintillation sheets. Besides, $A_{5,6}$ had a
hole in the middle and was seen by two PMs from two opposite corners
while $A_7,\ A_8$ were read by one PM attached to a corner of each.

All the counters were wrapped with aluminized
mylar.  Last
dinodes of all the PMs were powered from separate high voltage
sources to improve their stability.

\section{Magnetic spectrometer}
The wide-aperture magnetic spectrometer was based on the modified 
magnet SP-40~(M) with a uniform magnetic field in the volume of 
$100\times 70\times 150$~cm$^3$ and $p_T = 0.588$~GeV/$c$. The
spectrometer was equipped with the proportional chambers (PC), the drift tubes
(DT) and the hodoscopes $H_{4X}$, $H_{5X}$, $H_{6X}$, $H_{7}$, $H_{8Y}$.
The information from hodoscopes was used to generate trigger signals
and also to improve the performance of the track finding procedure.

\subsection{Proportional chambers}


The were five $X$- and five $Y$-planes arranged in
$z$-coordinate ascending order as follows:
$Y\!X\, Y\!X\ XY$\\$XYXY$. 
The design of the chambers is 
described in~\cite{pcs} with the exception that some dimensions were
slightly different. They as well the other PC system characteristics
are summarized in the Table~\ref{t:mwpc}.  

\begin{table}[hbt]
\caption{Basic characteristics of the PC chambers \label{t:mwpc}}
\center{
\begin{tabular}{|l|c|c|}  \hline
Parameter & X-chamber & Y-chamber \\ \hline \hline
Aperture (the inner size of frame)& $1315\times920$~mm$^2$ & $1315\times920$~mm$^2$
\\ \hline
Effective aperture (equipped with electronics) &
				     $768\times920$~mm$^2$ &
$1315\times640$~mm$^2$ \\ \hline

Wire pitch & 2 mm & 2 mm \\\hline
Number of channels (anode wires)  &  384 & 320 \\\hline
Anode wire diameter (gold-plated tungsten) & 20 $\mu$m & 20 $\mu$m  \\\hline
Cathodes (two aluminum foils)  & $2\times25\,\mu$m & $2\times25\,\mu$m \\\hline
Cathode--anode distance & 8 mm & 8 mm \\\hline
Number of additional wire supports & 1  & 2   \\ \hline
\end{tabular}}
\end{table}
%

Anode wire signals came to chamber-mounted shaping
preamplifiers (the shaping time $\sim 100$~ns, the sensitivity
threshold $\sim 3\,\mu$A) and then via 65~m long twisted pairs
arrived to 
latches (the gate $\sim 150$~ns). 
The gas mixture used for chambers was ${\rm Ar}\,(70\%) +
{\rm C}_4{\rm H}_{10}\,(27.8\%) + {\rm CF}_3{\rm Br}\,(0.2\%)+{\rm
C}_4{\rm H}_7{\rm OH}\,(2\%)$.
The overall efficiency of the chambers was found to be greater than 95\%.

\subsection{Drift tubes}

The DT system was  the 18 plane assembly. Each plane consisted of 32
thin-walled aluminum-mylar round tubes
with inner diameter of 62~mm. 
Each two consecutive planes formed a layer so that the
planes had the same orientation and a 10~mm shift along the sensitive axis.
The structure of the layers was as follows: 
$2X\,2S_1\,2X\,2S_2\,2X\,2S_1\,2X\,2S_2\,2X$, where
$X$ corresponded to vertically aligned tubes,
$S_{1,2}$, so-called ``stereo''-layers, had tubes 
inclined by $7.5^{\circ}$ and $-7.5^{\circ}$ 
from the vertical direction correspondingly. 
The distance between tube centers in the plane was 62.5~mm, the
distance between centers of planes in the layer was 64~mm, the
distance between centers of adjacent planes belonging to different
layers was also 64~mm.

\begin{figure}
\begin{minipage}{\pw}
\includegraphics[width=\pw]{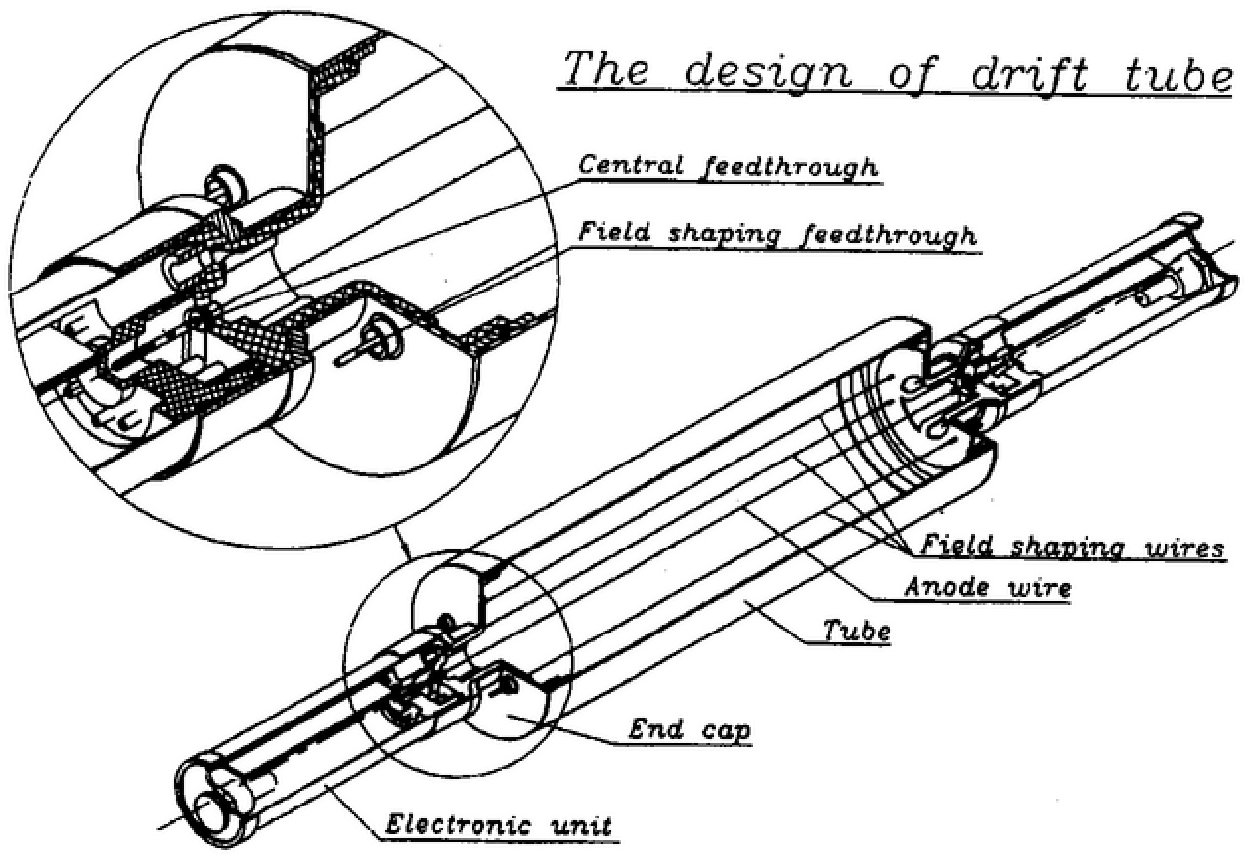}
\end{minipage}\hfill
\begin{minipage}{\pw}
\includegraphics[width=\pw]{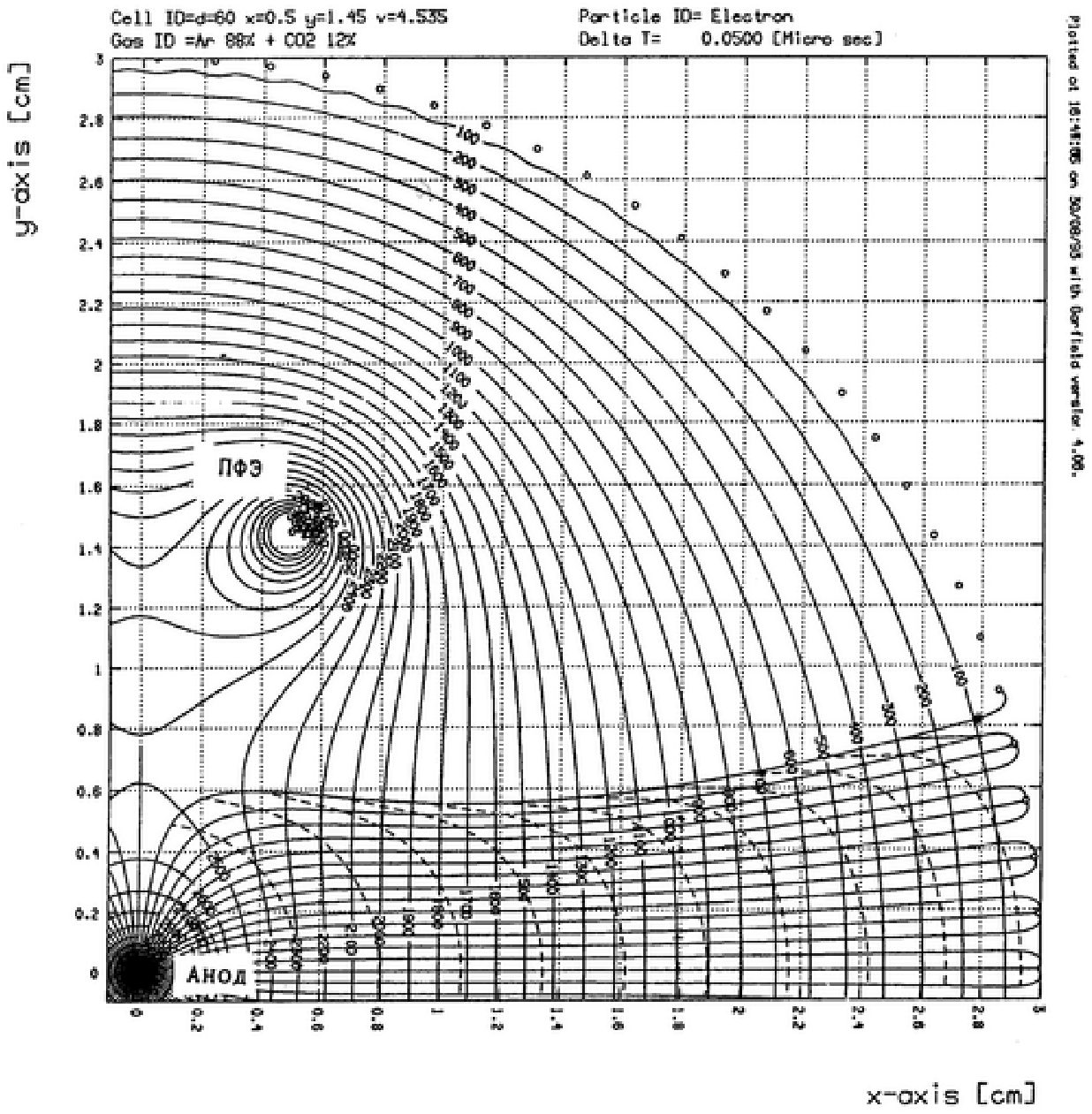}
\end{minipage}\hfill
\caption{ a) The design of the drift tube; 
b) Calculated electron field lines and equipotentials with a voltage at the anode and
the field-shaping wires of 4.5~kV.}
\label{fig01-dt}
\end{figure}

Each tube was made of $15\,\mu$m thick aluminum foil and
tree layers of $70\,\mu$m thick mylar~\cite{dt03} (see
Fig.~\ref{fig01-dt}a). 
It had one central anode wire
(gold-plated tungsten, $50\,\mu$m diameter) and four field-shaping
wires (stainless steel, $200\,\mu$m diameter). A voltage of $\sim
4$~kV was applied to all 5~wires, the tube walls were grounded and
served as cathodes. 
The field-shaping wires were needed to overcome the known drawback of ``classic'' DT
where the fast decrease of electric field strength (as $1/r$, where $r$ was
the distance from the anode wire) did not allow, in practice, to achieve the
electron drift velocity saturation for tubes with the diameter larger
than 30~mm. The field-shaping wires allowed to concentrate the electric field
lines coming from the anode wire in the narrow (about 10~mm wide) region
(Fig.~\ref{fig01-dt}b). The electric field strength sufficient for the
drift velocity saturation could be achieved that way. 

The geometry of the DT assembly was designed so that, if properly
positioned, a beam crossed each plane at approximately the center of
the 16-th tube. The sensitivity in these beam regions 
in such tubes was lowered by reducing  the electric field strength
there with a special electrode.
The electrode was a
5~cm aliminum foil and mylar cylinder with a diameter equal to the inner diameter of
the tube with a separate voltage applied to it. 
The gas mixture of ${\rm Ar}\,(88\%)+{\rm CO}_2\,(12\%)$ was used with
the tubes. Besides being 
inflammable, cheap and easily accessible this mixture is known to have a
property that the drift velocity saturation can be achieved 
at quite low electric field strength ($700$--$800$~V/cm).

The readout electronics included linear shaping preamplifiers
(tube mounted), multiplexors and time-digital converters
(TDC)~\cite{tdc}. The shaping time of the preamplifier was $\sim
80$~ns, the threshold sensitivity was 1.5~$\mu$A. The multiplexor had
40 paraphase $20\,\Omega$ input channels (only 32 were really used fed by
twisted pairs coming from the tubes of one plane) and 16 analog output channels
connected to TDCs. 
The multiplexors allowed  to record up to 4 time hits from 32 tube
plane using only 16 one-hit TDCs. 
For that inputs/outputs of the multiplexor were divided for four groups
12-4-4-12/4-4-4-4 so that the each group consisted of 12~side or
4~central tubes was served by 4 TDCs.
It worked as follows: the
tube numbers corresponding to the first hit in the group were memorized and the
first TDC channel was activated to measure its time, the second TDC channel was activated
in response to the second hit and so on. The smaller number of tubes in two
central groups accounted for the  heavier load at the central part of
the plane.
The time separation between hits in one group of tubes was $\sim
20$~ns. The TDC had a bin width 0.3~ns with integral nonlinearity
$\leq 0.2\%$ in the range up to 1000~ns.


The space resolution of the DT system was found to be $\approx
300\,\mu$m per plane (this also included uncertainties induced by
calibration, alignment, track model, etc.)

Further details on characteristics of the tubes and geometry
optimization procedure can be found
in~\cite{dt02} and~\cite{dt01} correspondingly.

\subsection{Downstream hodoscopes}

The $H_{4X}$ hodoscope located at the entrance to the magnet was made
of 10~mm thick scintillator manufactured in Kharkhov
(USSR). Dimensions of each of 24 elements seen by PMs from one side  were
$26\times400$~mm$^2$. 

The $H_{5X}$ hodoscope, located downstream of the magnet, was made of
10~mm thick extruded scintillator manufactured in IHEP. 
The hodoscope consisted of 48
elements of $48\times1000$~mm$^2$ seen by PMs from one side. 

The  $H_{6X}$ hodoscope situated after the DT system. It was also
made of 10~mm thick extruded IHEP manufactured scintillator. There
were 32 elements, $65\times1140$~mm$^2$ in size seen by PMs from one side.

The $H_7$ hodoscope, located just after the Cherenkov detector \v{C},
had a cell structure. The matrix $8\times 4$ of 20~mm thick
scintillation plates of the size $350\times 
300$~mm$^2$ was geometrically matched with $8\times 4$ mirror matrix
inside the Cherenkov detector. Such a system could be used for
trigger decisions. 
The scintillator was manufactured in Kharkov. Each plate was attached
to an individual PM.

Following $H_7$ the hodoscope $H_{8Y}$ 
physically consisted of two 
assemblies placed so that one corresponded to negative and the other one
to positive $x$-coordinates
with a small overlap (see Fig.~\ref{fig-sphinx21}). That allowed to
cover the aperture of 2.8~m in horizontal direction. Each assembly
consisted of 32 elements of  $48\times 1400$~mm$^2$ made of 10~mm thick 
extruded IHEP manufactured scintillator and seen by PMs from one
side. Thus the hodoscope could measure 32 different Y-coordinates and had 64 channels.

All cut surfaces of hodoscope elements were polished. Each element
was wrapped with aluminized mylar.
All the PMs were of FEU-85 type with the additional magnetic field
shielding consisted of multiple layers of Permalloy. Last dinodes of
the PMs attached to the central hodoscope elements which were located in the beam
region and worked at higher intensity were
powered from separate high voltage sources to improve their stability. Signals from PMs
were transported via 50~m long  $50\,\Omega$ coaxial cables to 
latches of registering electronics.

\section{Cherenkov counters}

The system of Cherenkov counters served for the identification of
secondary particles and included the RICH velocity spectrometer and
hodoscope-like threshold Cherenkov counter \v{C}.

\begin{figure}[htb]
\begin{center}
\leavevmode
\includegraphics[width=0.7\hsize]{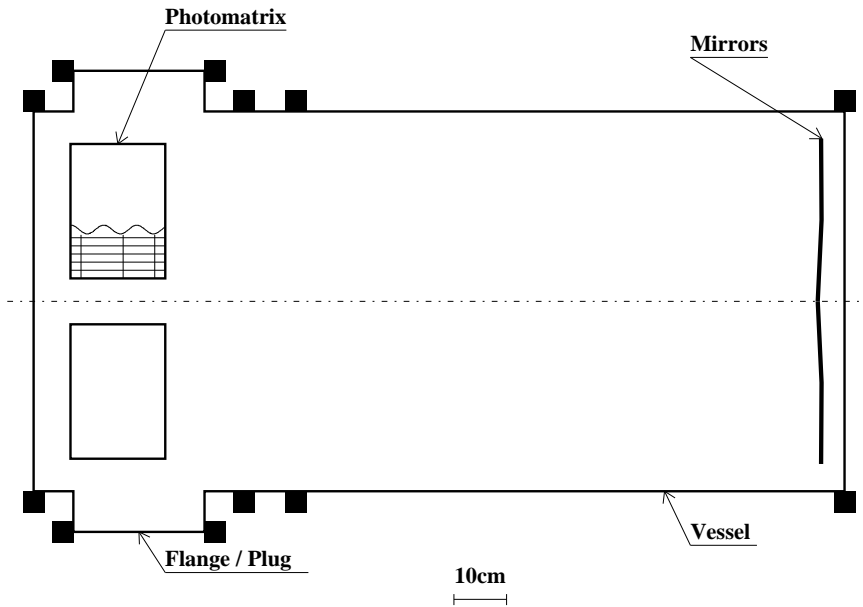} 
\end{center}
\caption{Schematic view of the RICH detector
(cut in horizontal plane).}
\label{rc1-dev1}
\end{figure}

Schematic view of the RICH detector is presented in Fig.~\ref{rc1-dev1}.
The vessel of the detector was made of steel,
$150\,{\rm cm}$ long and $70\,{\rm cm}$ in diameter.
It consisted of two parts to allow
easy access to photodetectors. Entrance and exit flanges were
made of $0.8\,{\rm mm}$ aluminum.

Sulfurhexafluoride~(${\rm SF}_6$), also known as elegas,
was used as a radiator. The working pressure was slightly
above atmospheric one (up to 10--15\%). The typical value
of the refractive index was 1.0008, what 
corresponded to the $\pi/K/p$ threshold momenta of $3.5/12.4/23.6\,{\rm GeV}/c$.

The optical system consisted of 2 identical spherical ($F=125\,{\rm cm}$)
glass mirrors of  rectangular ($30{\times}35\,{\rm cm}^2$) shape.
The important features of the detector were that
the mirrors were thin ($0.5\,{\rm cm}$) and  each mirror
focused light into one photomatrix, thus avoiding the need
for  precise alignment before the run.

\begin{figure}[htb]
\center{\includegraphics[width=0.7\textwidth]{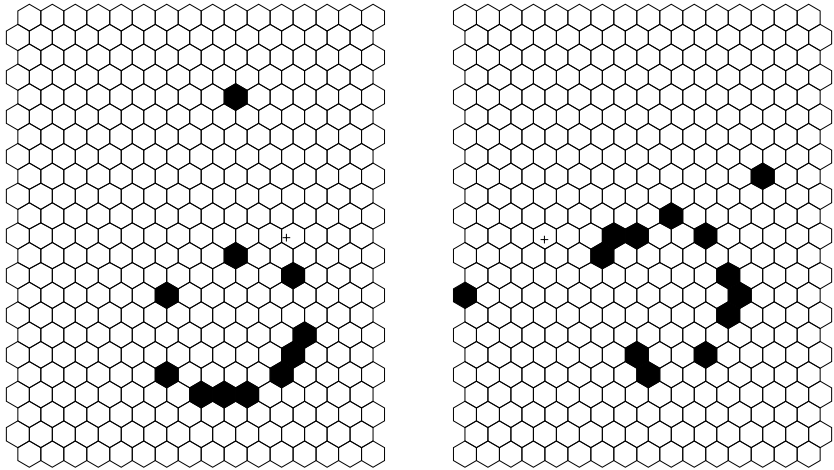}}
\caption{Layout of the RICH photodetector and example of an event.
Mirror alignment (reflection positions of the parallel
to detector axis light ray) is shown by the "+" markers.}
\label{rc1-sed}
\end{figure}

Photodetector consisted of 2 photomatrices,
each having 23 rows and 16 columns,
736 photo-multipliers in total.
Phototubes were hexagonally packed with $1.5\,{\rm cm}$ step
(see Fig.~\ref{rc1-sed} for layout).
Russian-made FEU-60 phototubes with
glass entrance window and $1.0\,{\rm cm}$ photocathode diameter were used.
To improve light registration efficiency
aluminized mylar cones and wavelength shifters (WLSs) 
were utilized.

WLSs were made according to the technology developed at IHEP. 
A thin ($20\,\mu m$) film was covered by the  $\simeq 5\,\mu m$ thick
transparent layer of  p-terphenyl.
Circle shape cutouts of such film based
WLSs were attached to the FEU-60 entrance windows with an optical
grease. 
The WLSs allowed to utilize the ultraviolet part of Cherenkov light
spectrum (which would be otherwise absorbed by glass windows) by shifting
it to wavelengths of maximum spectral sensitivity 
of the PMs. 
The number of Cherenkov photons emitted per unit wavelength
is  defined by the formula 
$dN/d\lambda \propto 1/\lambda^2$ which
shows that a significant
gain can be achieved by using the more intense shortwave part of the
spectrum. 
According to~\cite{ryk2,ryk3} 
the sensitivity of glass window PMs with
p-terphenyl based WLSs can be increased by a
factor of 2--3.5.
Measurements
carried out at IHEP with the $\mathrm{MgF}_2$ crystalline radiator,
a batch of FEU-60 PMs and the described WLSs showed the increase in
the number of photoelectrons by a factor of 2.5.


Phototubes had operating voltage in the 1--2$\,{\rm kV}$ range.
They were powered via resistor chains, which provided the possibility
of the voltage adjustment for each tube.
Preamplifiers with a sensitivity of 3--5$\,{\rm mV}/\mu{\rm A}$
were attached to the phototubes. Amplified signals were transmitted 
via $2\,{\rm m}$ long coaxial cables to the discriminators
with $10\,{\rm mV}$ threshold, which were located outside the vessel.
The discriminators generated differential ECL signals which were transmitted
with twisted pairs to the latches of custom-made registering
electronics.

The performance characteristics of RICH are given in
Table~\ref{t:rich} (see~\cite{rich2} for more details on the RICH).
\begin{table}[hbt]
\caption{The performance characteristics of RICH\label{t:rich}}
\center
\begin{tabular}{|p{8cm}|l|} \hline
Noise level (beam-on) & $\sim 1.5$ hits/trigger \\ \hline
Noise level (beam-off) & $\sim 1$ hits/trigger \\ \hline
Average number of hits per maximum radius ring  & 10.3
\\ \hline
$K$/$\pi$ separation at $2\sigma$ level up to momentum of & 30 GeV \\ \hline
\end{tabular}
\end{table}

The vessel of the multichannel counter \v{C} made of steel was $300$~cm long and 
$300$~cm in diameter. It housed 32 spherical mirrors of rectangular
shape of $350\times300$~mm$^2$  ($8\times 4$ matrix) focusing Cherenkov
light to the 32 PMs FEU-39. The counter was filled with air at atmospheric
pressure and had threshold momenta for $\pi/K/p$ equal to
6.0/21.3/40.1~GeV/$c$. The counter \v{C} together with the hodoscope $H_7$ (the
matrix of the same size) allowed to construct simple particle
identification requirements which could be used in the trigger.

\section{Electromagnetic calorimeter}

A multichannel lead glass electromagnetic calorimeter~(ECAL)
had been previously used in the EHS experiment at
CERN~\cite{gams} (known as Intermediate Gamma Detector there). 
It had a standard design of a two-dimensional lead glass matrix which
allowed to  measure both the position and the energy of a
shower. 

The initial assembly was rearranged to reduce the originally big
central hole to the size of one counter. The hole was still needed to
let the beam pass through.
The ECAL array of $39\times 27-1$ 
counters covers a surface of $1.95\times1.35$~m$^2$ 
(horizontal$\times$vertical). 
The individual counters with dimensions $5\times5\times 42$~cm$^3$   were
made of lead glass F8 manufactured in USSR, with radiation length $X_0=2.8$~cm. The
ECAL 
was mounted on a movable two-coordinate platform so that each counter
could be calibrated individually by exposing it to an electron beam.

The PMs (FEU-84) were shielded individually against the magnetic field
of the order of 20~G. The shielding consisted of cylinders of
``steel-50'' and Permalloy.  

A new monitoring system was built for ECAL consisting  of four LED
sources and a light 
distribution system which had light guides coming individually to the
front of each lead glass counter.

The typical energy and position resolutions obtained with an electron
beam were 
$\frac{\Delta E}{E} = \frac{0.15}{\sqrt{E/{\rm GeV}}} +0.02$~(FWHM) and
$\sigma \simeq \frac{8.2~{\rm mm}}{\sqrt{E/{\rm GeV}}}$
correspondingly.

\section{Hadron calorimeter}

A hodoscopic hadron calorimeter can be thought of as
a $12\times 8$ matrix of full absorption counters of
$20\times20$~cm$^2$. The calorimeter aperture was $2.5\times
1.6$~m$^2$, the total weight was 30~t.
Actually an independent element of the calorimeter was an assembled
still container housing four counters each consisted of 80 plates of
20~mm thick steel ($\approx 5$ nuclear interaction lengths) with
interlayers of plastic scintillator plates, 5~mm thick. The light was
collected by means of wavelength shifting  light-guides seen from
one side by PMs FEU-110.

The energy resolution for hadrons in the energy range 
$E=4$--$40$~GeV was $\sigma_E/E = 0.02+0.52/\sqrt{E}$ where $E$ is in
GeV and the mean
space resolution was $\approx 2$~cm at 40~GeV.

More information on the hadron calorimeter can be found in~\cite{ecal}.

\section{Trigger}

The two-level trigger scheme
was used with the SPHINX facility (to reduce DAQ system dead time). At zero level so called
strobe signals were generated. There were three different strobe
signals described by the logical formulae:
\begin{eqnarray}
T_0^b &=& S_1 S_2 S_3 S_4 \cdot (\overline{B_1 B_2}) \cdot
\overline{A_{5-8}}\cdot Gate \cdot \overline{Busy}, \\ 
T_0^c &=& S_1 S_2 S_3 S_4 \cdot (\overline{B_1 B_2}) \cdot
                         Gate \cdot \overline{Busy}, \\
T_0^d &=& S_1 S_2 S_3 S_4 \cdot Gate \cdot \overline{Busy} \cdot Prescale,
\end{eqnarray}
Their logical ``OR'' was one of the input signals for the next level trigger.
The $Gate$ was synchronized with the ``begin of spill'' signal,
the $Busy$  was used to prevent generating $T_0$ for the time the
decision was being made by the next level trigger and possible
subsequent readout cycle, the $Prescale$ was the signal from the prescaling
generator reducing the original rate of this strobe by a factor of
$\simeq 16000$.

The block of the first level trigger logic allowed to generate
up to eight signals using as primitive elements information from scintillation and veto
counters, multiplicity of hits in the hodoscopes and   
the threshold Cherenkov counter and total energy sum in the ECAL.
It also produced the logical ``OR'' of the above signals which served
as a ``general'' trigger and started event readout. 

The vast majority of the collected statistics corresponded to so
called three-track trigger
\begin{equation}
 T_{(3)} = T_0^b\cdot H_3(0{-}1)\cdot H_4(2{-}3)\cdot H_6({\equiv} 3)\cdot
H_7(1{-}3), \label{m:t3}
\end{equation}
where $H_i(m_1{-}m_2)$ means multiplicity requirement between $m_1$ and
$m_2$ in hodoscope $H_i$.
It allowed to select events with three charged tracks in
the final state 
produced in exclusive reactions. 
Other important triggers used during data taking runs are listed
below:
\begin{equation}
\label{m:t5}
T_{(5)} = T_0^b\cdot H_3(0{-}1)\cdot H_4(4{-}5)\cdot
H_5({\geqslant}4)\cdot H_6({\equiv} 5)\cdot 
H_7(3{-}6)
\end{equation} 
or five-track trigger allowed to select exclusive reactions with five secondary
tracks;

\begin{equation}
\begin{array}{rcl}
T_{(2)} &= T_0^c\cdot(&\!H_3({\geqslant} 1)\cdot H_4(1{-}2)\cdot H_5(2{-}3)
                     \cdot H_6({\equiv} 2)\cdot H_7({\equiv} 2)\ +\\ 
		     &&\!H_3({\geqslant} 1)\cdot H_4(2{-}3)\cdot H_5(2{-}4)
                     \cdot H_6({\equiv} 3)\cdot H_7(2{-}3)\ ) \label{m:t2}
\end{array}
\end{equation} 
or ``meson'' trigger selected inclusive reactions with two or tree charged tracks in
the final state;

\begin{equation}
\label{m:t7}
T_{(neu)} = T_0^c\cdot H_3({\geqslant} 1)\cdot H_5({\equiv}0)\cdot H_6({\equiv} 0)\cdot
H_7({\equiv}0)\cdot E_{\rm ECAL}>0
\end{equation} 
or neutral trigger
selected events with only neutral secondary particles and some energy
deposited in the ECAL ($E_{\rm ECAL}>0$);

\begin{equation}
\label{m:t4}
T_{(beam)} = T_0^d
\end{equation}
or beam trigger --- prescaled trigger on a beam particle.

All these trigger signals could be used simultaneously. In
reality, different subsets of them were used during different periods
of data taking runs.
This allowed to select many exclusive
reactions and calibration 
processes, part of which will be discussed in section~\ref{se:illu}.


\section{DAQ hardware}

DAQ architecture was based on IHEP developed MISS
standard~\cite{miss0}. Some modules were in IHEP SUMMA standard~\cite{summa}.
The system topology had four-level structure:
%
\begin{enumerate}
\item
{\bf Front-end electronics level} (lowest one) included modules of latches,
TDCs and ADCs as
well as crate controllers and special module controllers. All modules
were of either  MISS or SUMMA types.

\item {\bf Branch buffer memory level}
served as an intermediate storage of event information from a single branch and for
partial event building. All modules
were of MISS type. 

\item {\bf System buffer memory level} served as an intermediate storage
of event information from the whole system and for event building. Its design was similar to
the branch buffer memory level.

\item {\bf Data accumulation level} (spill data level)
built a packet containing information on all events happened during
one spill. In addition, apparatus control and
monitoring functions were performed at this level.
\end{enumerate}

Table~\ref{t:ch} gives the numbers of channels used to record physical
information for the SPHINX setup  arranged by module type.
\begin{table}[hbt] \center
\caption{Channels of SPHINX setup arranged by the front-end module type \label{t:ch}}
\begin{tabular}{|l|p{7cm}|r|} \hline
Type & Detector & Number of channels \\ \hline \hline
ADC & HCAL, ECAL & $\sim$  $1400$ \\ \hline
TDC & DT & $\sim$ $\,\ 700$ \\ \hline
Latches & PC, hodoscopes, RICH & $\sim$ $5000$\\ \hline\hline
Total && $\sim$ $7100$ \\ \hline
\end{tabular}
\end{table}
All inter-level connections were also designed according
to MISS specifications. More information on the electronics used in the
system can be found in~\cite{miss,miss2,miss3,miss4}.

\begin{figure}[hbt]
\center{ 
\includegraphics[width=\textwidth]{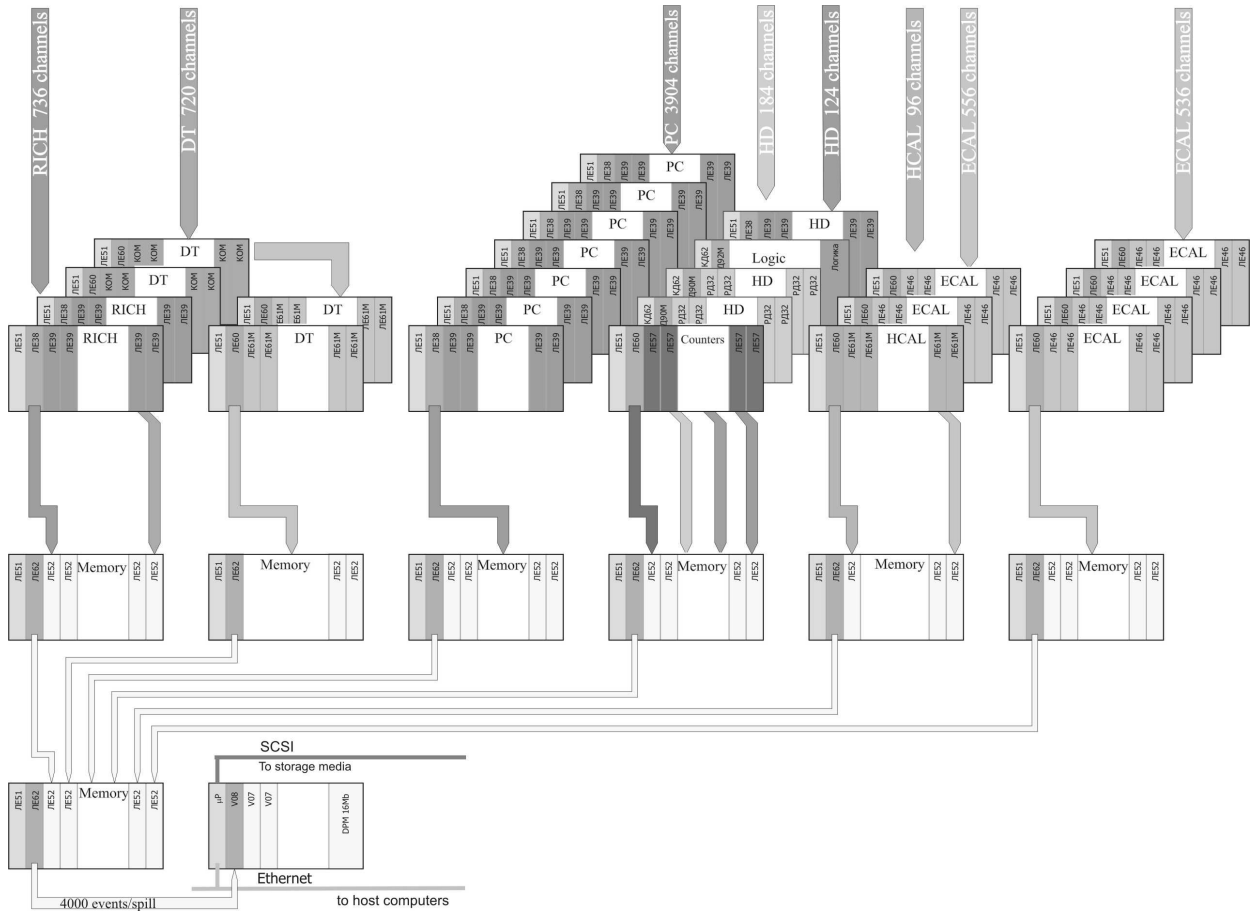}}
\caption{Scheme of the DAQ of the SPHINX setup\label{f:daq}}
\end{figure}

The general scheme of the SPHINX DAQ system is shown in Fig.~\ref{f:daq}.
The lowest (first) level had 25 sectors\footnote{the MISS sector is a
group of 8--24 modules sharing the same ECL bus, see~\cite{miss0} for details.} 
of registering electronics. 
Each
sector of the second level was
connected to $2$--$9$ first level sectors to keep data flow at the
optimum level. 
The system could use
SUMMA crates as MISS sectors at the first level. Additional
special controllers in such crates simulated  standard
MISS controllers. These SUMMA crates contained modules of a trigger
logic.

The working algorithm was fully based on standard MISS procedures. The
``begin of spill'' signal 
started a reset cycle. 
Event selection was done in two stages with generating of strobe signals
and trigger signals at the first and the second stage correspondingly.
A set of strobe signals was produced
by a strobe-block and an auxiliary logic fed by information from the beam
counters and the  guard system. The logical ``OR'' of these signals
allowed registering information from hodoscopes, PC and RICH (all latches). 
It also put the strobe-block into a busy state preventing it from
generating further strobes. 
The information on strobe type, multiplicity in the particular set of
hodoscopes and on energy deposited in the ECAL came to a trigger block~\cite{trigg}
where the final decision 
on whether to accept an event or not was made. If it was
negative, the fast reset signal was distributed to the activated part
of the electronics followed by clearing the strobe-block busy state. 
If positive, the information
from the ECAL, the HCAL and the DT was registered. This signal also initiated
the process of the front-end electronics readout and moving the data
to the next level. This was accompanied by extending the strobe/trigger
logic busy state for the time of the readout.
When done the busy state was cleared and the
system was ready for further data taking. As the data collection
proceeded, the information moved up through the intermediate levels and
was accumulated at the top level. 

Thus, given that the  ``begin of spill'' signal was provided, the hardware part
of the DAQ system was capable to collect, build and store the
information on the events collected during the spill according to
trigger types programmed 
into the trigger block without any outside intervention. 
This information was stored in the 16~MB buffer memory
module of the top level VME crate.

The ``begin of spill'' signal was provided by a software. The software also generated
additional trigger signals of special types for 
calibration data collection
and saved all the collected data  to a permanent
storage media.

Some basic DAQ
characteristics are given in the Table~\ref{t:daq}.
\begin{table}[htb]
\caption{Characteristics of the SPHINX DAQ system\label{t:daq}}
\centerline{
\begin{tabular}{|l|l|} \hline
spill/cycle time & 1.5 sec/9 sec \\ \hline
trigger rate & 4 KHz \\ \hline
event length  & 1.3 KByte \\\hline
data flow & 5.2 MByte/sec \\ \hline
live time & $\sim 70\%$ \\ \hline
\end{tabular} }

\end{table}

\section{DAQ and online software}

The DAQ software provided the following functions: 
start/stop of the basic data collection mode (which includes
calibration data collection),
writing the collected data to disks and/or
magnetic tapes,
online apparatus monitoring,
initiating special modes for apparatus and DAQ tests and calibrations.


The DAQ software worked mostly under LynxOS operating system installed on
a one-board MVME-167 computer with Motorola-68040 CPU and 16~MB of
memory. The computer was located in the top level VME crate. The top level
memory buffer (also 16~MB) located in the same crate was mapped
to the computer memory address space so that it could be accessed by the
processor as a part of its own memory.

The software design allowed to
distribute a
part of the event flow (including the whole) to consumers on other
computers via 
Ethernet. The online monitoring worked this way so that apparatus control
parameters could be analyzed with two workstation (DECStation 5200 and
Silicon Graphics Indy) working under UNIX-type OS.

The software had the following modules (programs):
\begin{enumerate}

\item a dispatcher managing the whole system;
\item data collection modules;
\item disks/tapes writing modules;
\item data distributor;
\item DAQ test and control modules;
\item apparatus test and calibration modules.

\end{enumerate}
The dispatcher was the core of the system. It accepted messages from
the connecting modules and communicated them to other modules
according to the type of the sending module and the message content 
for actions.

The data collection modules consisted of the main program
working on in-spill information and auxiliary programs which were not
necessarily spill-tied. The main program activated by the ``begin of
spill'' signal (coming from an
accelerator control apparatus), initialized the DAQ apparatus and
initiated three special
sequences: for pedestal measurements, for collecting data with the ECAL and
the RICH monitoring systems, for collecting TDC calibration data. By the ``end
of spill'' signal it also sent the message with  parameters of the collected data buffer 
to the dispatcher thus making
this information available for other modules. Another data collection
module read out scalers. 
The reading sequence started
by the end of the spill. The collected information then was
communicated (through the dispatcher) for writing to disks and for
displaying the scalers 
readings.

The data could be written to disk files which could be
then written to tapes or directly to tapes (an Exabyte drive
was attached to the MVME-167). 

There was the distributing module which could send out the information
on the  uniform subset of 
events via Ethernet. This was used for the online apparatus monitoring. On the
other side (workstations) the information was further distributed to
different analyzing processes (analyzers). It was done in the framework of
UNIDAQ system~\cite{unidaq1,unidaq2}. One of the advantages of this system
was that analyzers were different programs which had a standard piece
of code allowing them to connect  to the central distributing unit (NOVA)
and got the data. Thus they could be written by different people for
different purposes. There were analyzers for filling and displaying
control histogram, for single event display, for online chamber
efficiency measurements with tracks, for testing the structure of
events built by electronics, for monitoring recoverable
electronic failures. Another special analyzer
monitored some, presumably constant, setup parameters and produced an
information message on a main console accompanied by an alarm signal
if these parameters came out of predefined windows.

The DAQ system could be controlled by sending messages to different
modules or to all modules at once with the special
program. There was a set of programs for the DAQ hardware testing and debugging 
and also for the apparatus testing and calibrations.

\section{Illustration of the experimental possibilities\label{se:illu}}

In this section we will consider very briefly several possibilities for
the measurements with the SPHINX facility with different triggers.

\subsection*{$\bf T_{(3)}$ trigger}

The three-track trigger $T_{(3)}$~(\ref{m:t3}) was the main trigger
for the measurement with the SPHINX spectrometer. It
allowed to study simultaneously the exclusive reactions:
\begin{eqnarray}
p+N({\rm C}) &\to & [p K^0_L] K^0_S+N({\rm C});\ K^0_S\to\pi^+\pi^- \label{m:9} \\
       &\to & [n K^+] K^0_S  +N({\rm C});\ K^0_S\to\pi^+\pi^- \label{m:10} \\
	&\to & p K^+ K^-  +N({\rm C}) \label{m:11} \\
	&\to & p K^0_S K^0_L+N({\rm C});\ K^0_S\to\pi^+\pi^- \label{m:12} \\
	&\to & p \phi+N({\rm C});\ \phi\to K^+ K^-,\ K^0_S K^0_L \label{m:13} \\
	&\to & \Lambda K^+ +N({\rm C});\ \Lambda\to p\pi^- \label{m:14} \\
	&\to & [\Sigma^0 K^+] +N({\rm C});\ \Sigma^0\to\Lambda\gamma;\ 
\Lambda\to p\pi^- \label{m:15} \\
	&\to & [\Sigma^+ K^0_S] +N({\rm C});\ K^0_S\to\pi^+\pi^-;\
\Sigma^+\to p\pi^0,\ n\pi^+ \label{m:16}  \\
	&\to & \Lambda(1520) K^+ +N({\rm C});\ \Lambda(1520)\to p K^-,\
\Lambda\gamma;\ \Lambda\to p\pi^- \label{m:17} 
\end{eqnarray}
and many other processes (reactions on quasi-free nucleons $N$ or
coherent reactions on $C$ nuclei). The long lived neutral particles
($n$, $K^0_L$, $\gamma$) were detected by their interactions in the ECAL
calorimeter and were not used in the trigger requirements.

\subsection*{$\bf T_{(5)}$ trigger}

The five-track trigger $T_{(5)}$~(\ref{m:t5}) was used, for example,
for the selection of reactions:
\begin{eqnarray}
p+N({\rm C}) &\to & [p K^0_S] K^0_S+N({\rm C});\ K^0_S\to\pi^+\pi^- \label{m:18} \\
	&\to & \Lambda(1520) K^+ +N({\rm C});\
\Lambda(1520)\to\Lambda\pi^+\pi^-;\ \Lambda\to p\pi^-\label{m:19} 
\end{eqnarray}

\subsection*{$\bf T_{(2)}$ and $\bf T_{(neu)}$ triggers}

These triggers (\ref{m:t2}) and (\ref{m:t7}) were used for the study
of quasi-exclusive forward meson production in proton reactions in deep
fragmentation region (inclusive for bottom vertex)~\cite{13}:
\begin{equation}
p+N\to (meson)|_{\rm forward}+(X)|_{\rm bottom\ vertex}
\end{equation}
These deep fragmentation processes can be accompanied by substantial
quark rearrangement, gluon forward bremsstrahlung and open new
possibilities for the search for exotic meson production~\cite{37}.

\vspace*{2em}

\begin{figure}[p]
\begin{minipage}[t]{0.45\hsize} 

\includegraphics[width=\hsize]{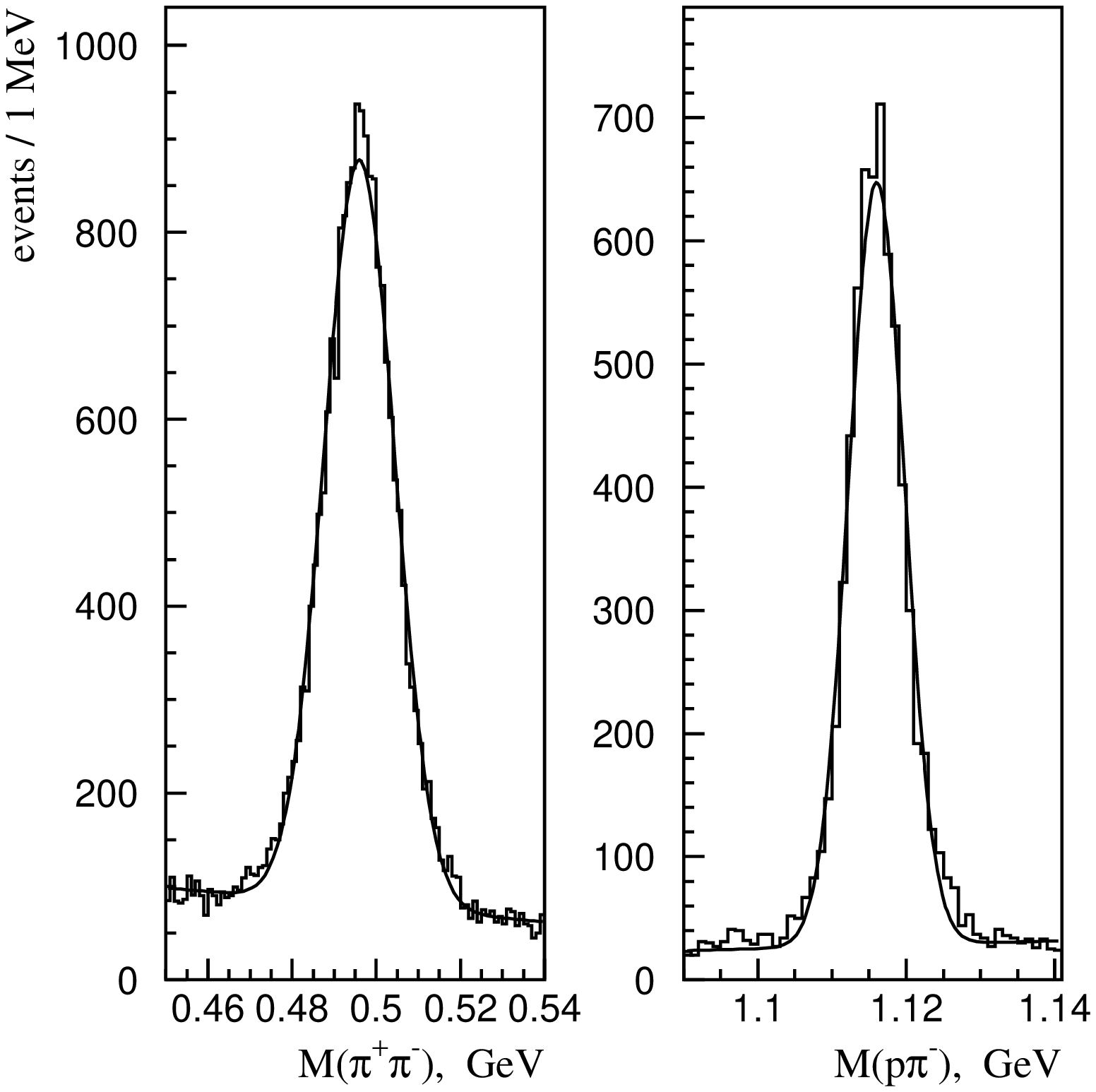}
\caption{%
Signals from decays of $K_S^0\to\pi^+\pi^-$ and
$\Lambda\to p\pi^-$ in the reactions $p + N \to nK_S^0K^+ + N$
and $p + N \to \Lambda\pi^+\pi^-K^+ + N$. The fit
gives: $M(K_S^0) = 496$~MeV, $\sigma(K_S^0) = 8.4$~MeV, 
$M(\Lambda) = 1116$~MeV, $\sigma(\Lambda) = 3.8$~MeV (see~\cite{17}). 
\label{f:6}}

\end{minipage}
\hfill
\begin{minipage}[t]{0.45\hsize}
\includegraphics[width=\hsize]{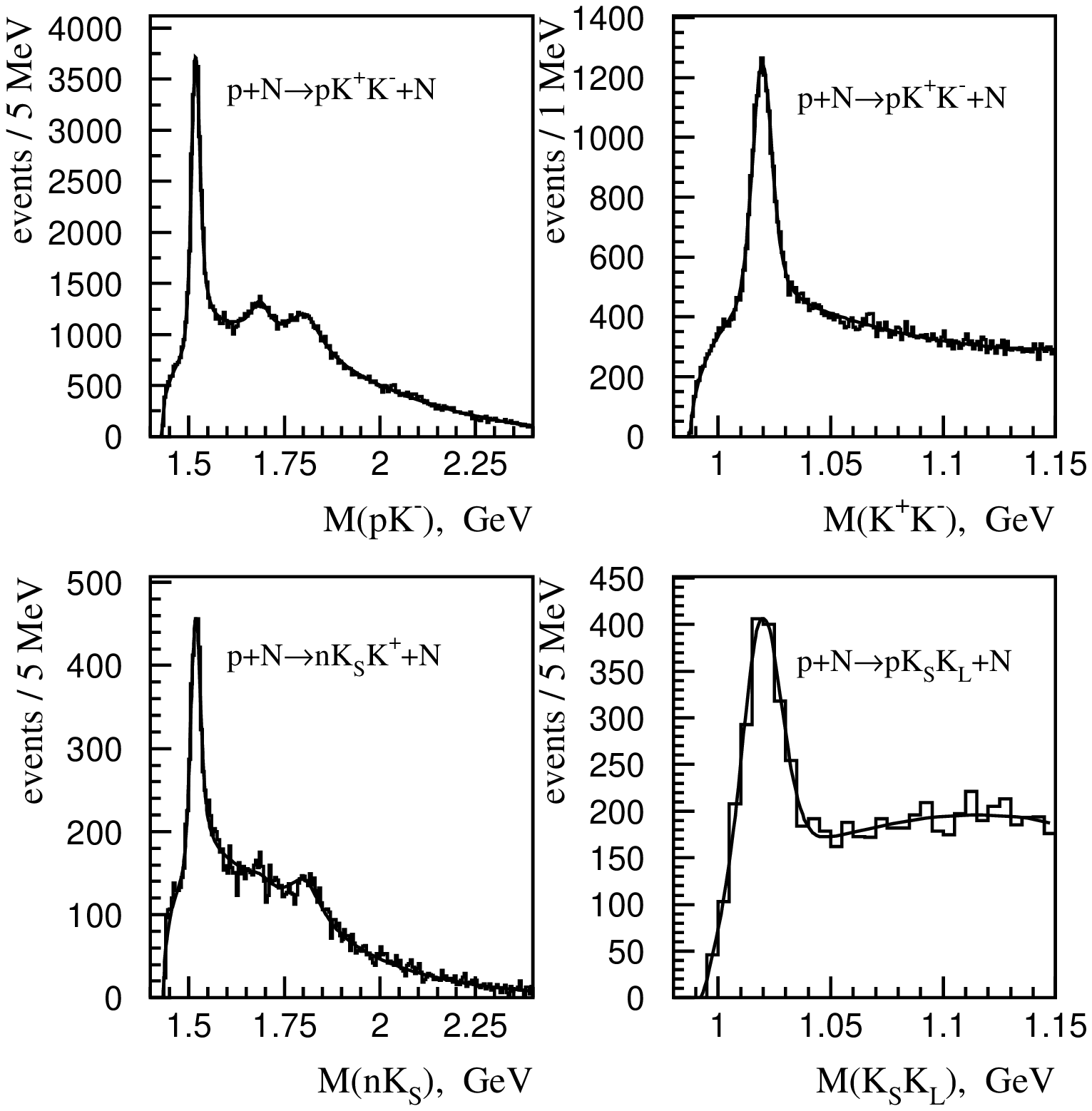}
\caption{
Effective mass distributions $M(pK^-)$, $M(nK_S^0)$ for the reaction
$p+N \to [N\bar{K}]K^+ +N$ (two left pictures), 
and  $M(K^+K^-)$, $M(K_S^0K_L^0)$ for the reaction
$p+N \to p[K\bar{K}] +N$ (two right ones; see~\cite{17}). 
\label{f:8}}

\end{minipage}
\end{figure}

The possibilities to select some of the above reactions are illustrated in
Fig.~\ref{f:6}--\ref{f:8} reproduced from~\cite{17}. Reactions (\ref{m:9}),
(\ref{m:10}), (\ref{m:18}) were used for the search for the exotic
pentaquark baryon $\Theta(1540)^+$ with positive strangeness and 
$uudd\bar s$ quark 
structure in the mass spectra of $[n K^+]$,
$[p K^0_S]$, $[p K^0_L]$ systems~\cite{17}. Reactions (\ref{m:15}),
(\ref{m:16}) were used for the search for the cryptoexotic $X(2000)$
baryon with hidden strangeness ($uuds\bar s$) in the mass spectra
$M(\Sigma^0 K^+)$ and $M(\Sigma^+ K^0_S)$~\cite{5,8,10,11,14,15,16}.

\begin{figure}[p]
\begin{minipage}[t]{0.45\hsize}

\includegraphics[width=\hsize]{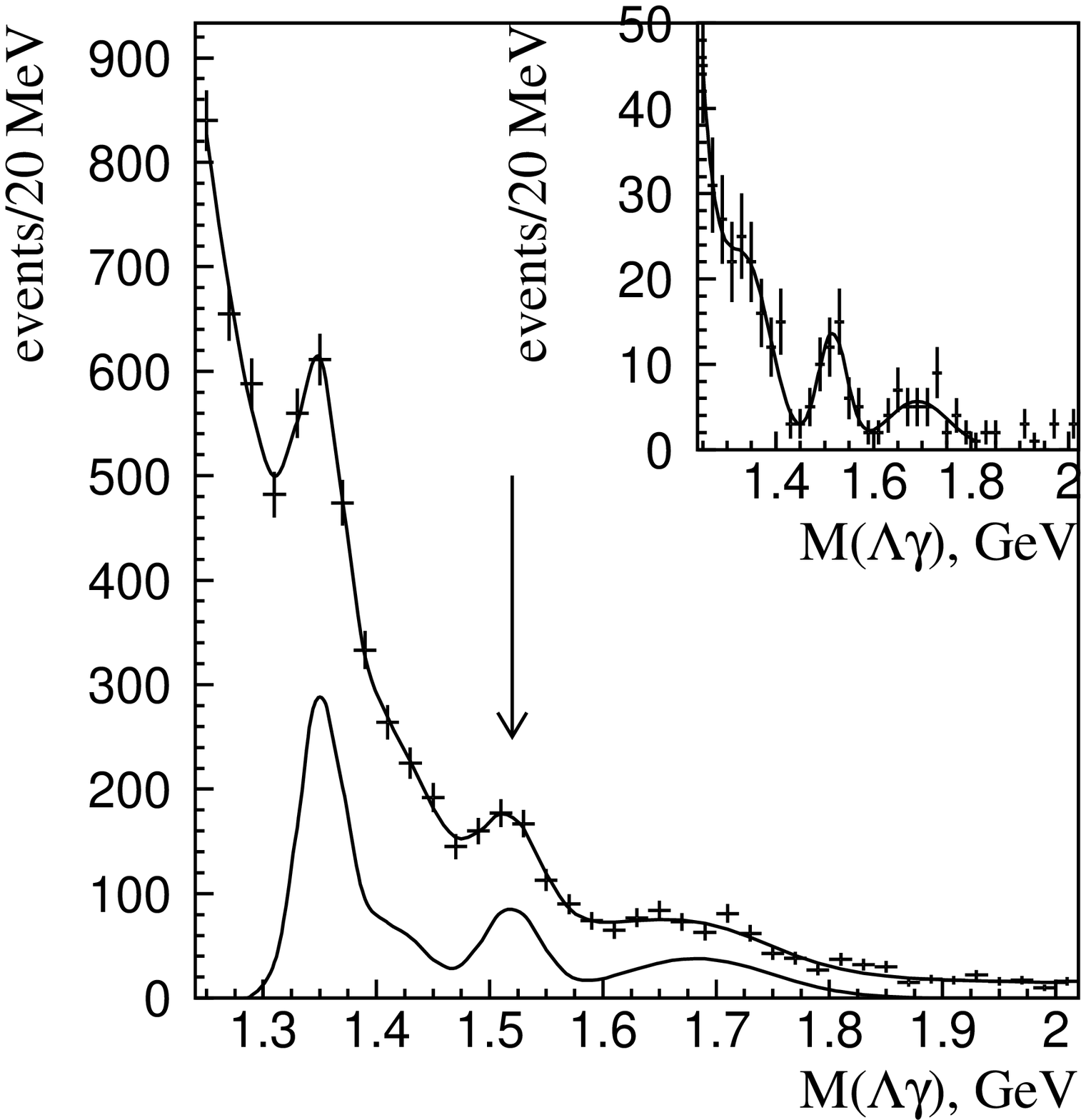}
\caption{%
Effective mass spectrum $M(\Lambda\gamma)$ in reaction
$p+N({\rm C})\to[\Lambda\gamma]K^+ +N({\rm C});\ \Lambda\to p\pi^-$. The arrow
shows the $\Lambda(1520)$ peak. The right-top picture presents the
same spectrum obtained with more stringent criteria for suppression of
extra photons~\cite{18}.
\label{f:11}}

\end{minipage}

\end{figure}

The SPHINX facility had the good guard system allowing to suppress
background processes with ``lost photons''. That opened the possibility
to study the rare radiative decay $\Lambda(1520)\to \Lambda\gamma$ in
the reaction (\ref{m:17}) --- see Fig.~\ref{f:11} reproduced from~\cite{18}.

\section{Conclusion}


The multipurpose wide-aperture SPHINX spectrometer, in its several
modifications, was 
used in the experiments at the proton
beam of IHEP U-70 accelerator in 1989--1999~[1--18]. The good quality
of this setup with the possibility of the complete reconstruction of
charged and neutral secondary particles and large collected statistics
allowed to get many results on the study of
diffractive production reactions, searches for new states of hadronic
matter, study of the radiative decays of hyperons, OZI rule and deep
fragmentation reactions~[1--18] and opened new perspectives for further
investigations of these and other rare processes.

\section*{Acknowledgements}

This work was partly supported by Russian
Foundation for Basic Researches 
(grants 99-02-18252, 02-02-16086 and~05-02-16924).



\begin{thebibliography}{99}

\bibitem{3} D. V. Vavilov {\it et al.} (SPHINX Collab.), Yad. Fiz. {\bf 57},
241 (1994) [Phys. At. Nucl. (Engl. Transl.) {\bf 57}, 227 (1994)]; 
M. Ya. Balatz {\it et al.} (SPHINX Collab.), Z. Phys. C {\bf 61}, 220 (1994).
\bibitem{4} M. Ya. Balatz {\it et al.} (SPHINX Collab.), Z. Phys. C {\bf 61}, 
399 (1994).
\bibitem{5} D. V. Vavilov {\it et al.} (SPHINX Collab.), Yad. Fiz. {\bf 57},
253 (1994) [Phys. At. Nucl. (Engl. Transl.) {\bf 57}, 238 (1994)].
\bibitem{6} L. G. Landsberg {\it et al.} (SPHINX Collab.), Nuov. Cim. A 
{\bf 107}, 2441 (1994).
\bibitem{7} D. V. Vavilov {\it et al.} (SPHINX Collab.), Yad. Fiz. {\bf 57},
1449 (1994); {\bf 58}, 1426 (1995) [Phys. At. Nucl. (Engl. Transl.) 
{\bf 57}, 1376 (1994); {\bf 58}, 1342 (1995)].
\bibitem{8} S. V. Golovkin {\it et al.} (SPHINX Collab.), Z. Phys. C {\bf 68},
585 (1995).
\bibitem{9} S. V. Golovkin {\it et al.} (SPHINX Collab.), Yad. Phys. {\bf 59},
1395 (1996)  [Phys. At. Nucl. (Engl. Transl.) {\bf 59}, 1336 (1996)].
\bibitem{10} V. A. Bezzubov {\it et al.} (SPHINX Collab.), Yad. Phys. {\bf 59},
2199 (1996) [Phys. At. Nucl. (Engl. Transl.) {\bf 59}, 2117 (1996)]. 
\bibitem{11} L. G. Landsberg, {\it Hadron Spectroscopy ("Hadron 97"). Seventh 
Intern. Conf. Upton, NY, Aug. 1997} (Ed. S.-U. Chung, H. J. Willutzki), p. 725.
\bibitem{12} V. A. Victorov {\it et al.}, Yad. Fiz. {\bf 59}, 1229 (1996) 
[Phys. At. Nucl. (Engl. Transl.) {\bf 59}, 1175 (1996)]; 
M. Ya. Balatz {\it et al.} (SPHINX Collab.), Yad. Fiz. {\bf 59}, 1242 (1996)
 [Phys. At. Nucl. (Engl. Transl.) {\bf 59}, 1186 (1996)];   
S. V. Golovkin {\it et al.} (SPHINX Collab.), Z. Phys. A {\bf 359}, 435 (1997).
\bibitem{13} S. V. Golovkin {\it et al.} (SPHINX Collab.), Z. Phys. A 
{\bf 359}, 327 (1997). 
\bibitem{14} S. V. Golovkin {\it et al.} (SPHINX Collab.), Eur. Phys. J. 
  A {\bf 5}, 409 (1999).
\bibitem{15} D. V. Vavilov {\it et al.} (SPHINX Collab.), Yad. Fiz. {\bf 63},
1469 (2000).
\bibitem{2} L. G. Landsberg, Phys. Rep. {\bf 320}, 223 (1999); 
L. G. Landsberg, Yad. Fiz. {\bf 62}, 2167 (1999).
\bibitem{1} L. G. Landsberg, UFN {\bf 164}, 1129 (1994) [Physics Uspekhi 
(Engl. Transl.) {\bf 37}, 1043 (1994)]; 
V. F. Kurshetsov, L. G. Landsberg, Yad. Fiz. {\bf 57}, 2030 (1994) 
  [Phys. At. Nucl. (Engl. Transl.) {\bf 57}, 1954 (1994)]; 
L. G. Landsberg, Yad. Fiz. {\bf 60}, 1541 (1997)  [Phys. At. Nucl. (Engl. 
Transl.) {\bf 60}, 1397 (1997)].

\bibitem{16} Yu. M. Antipov {\it et al.} (SPHINX Collab.),
Yad. Fiz. {\bf 65}, 2131 (2002) [Phys. At. Nucl. (Engl. Transl.) 
{\bf 65}, 2070 (2002)].

\bibitem{17} Yu. M. Antipov {\it et al.} (SPHINX Collab.), Eur.~Phys.~J.~A
{\bf 21}, 455 (2004);
V.F. Kurshetsov {\it et al.} (SPHINX Collab.), Yad.~Fiz. {\bf 68}, 468 (2005)
[Phys.~At.~Nucl. (Engl. Transl.) {\bf 68}, 439 (2005)].

\bibitem{18} Yu. M. Antipov {\it et al.} (SPHINX Collab.), Phys.Lett.~B
{\bf 604}, 22 (2004); 
D. V. Vavilov {\it et al.} (SPHINX Collab.), 
Yad.~Fiz. {\bf 68}, 407 (2005) [Phys.~At.~Nucl. (Engl. Transl.) {\bf 68}, 378 (2005)].

\bibitem{chan} A.A. Batalov {\it et al.}, Preprint IFVE-87-116 (1987) (in Russian).

\bibitem{pcs} A.V. Vishnevsky {\it et al.}, Prib.~Tekh.~Eksp. {\bf 1}, 60
(1984) (in Russian).

\bibitem{dt03} Yu. Antipov {\it et al.}, Nucl. Instr. and Meth. in Phys.
               Res.~A {\bf 379}, 434 (1996).

\bibitem{tdc} S.~Zimin, M.~Soldatov, Preprint IFVE-93-50 (1993) (in Russian).

\bibitem{dt02} Yu. Antipov {\it et al.}, Nuclear Physics B (Proc. Suppl.)
               {\bf 44}, 206 (1995).

\bibitem{dt01} Yu. Antipov {\it et al.}, Preprint IFVE-94-128 (1994)
(in Russian).


\bibitem{ryk2} A.M. Gorin {\it et al.}, Nucl. Instr. Meth.~A {\bf
251}, 461 (1986).

\bibitem{ryk3} P. Baillon {\it et al.}, Nucl. Instr. Meth. {\bf
126}, 13 (1975).


\bibitem{rich2}{
A. Kozhevnikov, V. Kubarovsky, V.~Molchanov, V.~Rykalin and V.~Solyanik, 
Nucl. Instr. Meth.~A {\bf 433}, 164 (1999).
}

\bibitem{gams} B. Powell {\it et al.}, Nucl. Instr. Meth. {\bf 198},
217 (1982). 



\bibitem{ecal} Y.M. Antipov et al., Nucl. Instr. Meth.~A {\bf 295},
81 (1990).

\bibitem{miss0} S.I. Bityukov {\it et al.}, Preprint IFVE-94-101 (1994) (in Russian).

\bibitem{summa} O.I. Alferova {\it et al.},  Prib.~Tekh.~Eksp. {\bf 4},
 56 (1975) (in Russian).



\bibitem{miss}   Yu.~Bushnin {\it et al.}, Preprint IFVE-88-47 (1988) (in Russian).
\bibitem{miss2}  Yu.~Bushnin {\it et al.}, Preprint IFVE-93-72 (1993) (in Russian).
\bibitem{miss3}  Yu.~Bushnin {\it et al.}, Preprint IFVE-95-88 (1995) (in Russian).
\bibitem{miss4}  Yu.~Bushnin {\it et al.}, Preprint IFVE-95-104 (1995) (in Russian).

\bibitem{trigg} A.~Kozhevnikov {\it et al.}, Preprint IFVE-91-101 (1991) (in Russian).

\bibitem{unidaq1}
M.~Nomachi {\it et al.} 
Proceedings of the International Conference on  Computing in
High-energy Physics (CHEP 94), LBL-35822, 114 (1994).

\bibitem{unidaq2} UNIDAQ collaboration, SDC-93-573 (1993); UMHE-93-29
(1993).

\bibitem{37} L.G. Landsberg, Yad.~Fiz. {\bf 52}, 192 (1990); 
Nucl.~Phys. (Proc.~Suppl.) {\bf 211}, 179 (1991).




\end{thebibliography}
\end{document}